\documentclass[apl, twocolumn,tightenlines,superscriptaddress]{revtex4-1}

\usepackage{amsthm}
\usepackage{amsmath}
\usepackage{bbm, dsfont}
\usepackage{graphicx}
\usepackage[usenames,dvipsnames]{color}
\usepackage[colorlinks=true,citecolor=magenta,urlcolor=blue]{hyperref}
\usepackage[table]{xcolor}
\usepackage{colortbl}
\usepackage{color}
\usepackage{multirow}
\usepackage{hhline}
\usepackage{tabu}
\usepackage{epstopdf}
\usepackage{booktabs}
\usepackage[normalem]{ulem}
\usepackage{siunitx}
\usepackage{enumitem}
\usepackage{braket}
\usepackage{siunitx}
\usepackage{textcomp}
\usepackage{booktabs}
\usepackage[textwidth=3cm, textsize=footnotesize]{todonotes}
\usepackage[font=small,skip=3pt]{caption}
\usepackage{bbold}
\usepackage[percent]{overpic}
\usepackage{makecell}
\usepackage{multirow}
\usepackage{float}
\usepackage{textcomp}

\DeclareSIUnit\permille{\text{\textperthousand}}

\DeclareSIUnit{\belmilliwatt}{Bm}
\DeclareSIUnit{\dBm}{\deci\belmilliwatt}
\DeclareSIUnit{\bps}{bps}
\DeclareSIUnit{\bit}{b}

\graphicspath{{./Figures/}}

\newcolumntype{C}[1]{ >{ \centering\arraybackslash}p{#1}}

\definecolor{gray4}{gray}{0.8}
\definecolor{gray2}{gray}{0.6}

\definecolor{red_plot}{rgb}{1.0, 0.0, 0.0}
\definecolor{green_plot}{rgb}{0.486, 0.988, 0.0}
\definecolor{cyan_plot}{rgb}{0.678, 0.847, 0.902}

\begin{document}
\title{The limits of multiplexing quantum and classical channels: Case study of a 2.5~GHz discrete variable quantum key distribution system} 

\author{Fadri Gr\"unenfelder}\email{fadri.gruenenfelder@unige.ch}
\affiliation{Group of Applied Physics, University of Geneva, Chemin de Pinchat 22, CH-1211 Geneva 4, Switzerland}
\author{Rebecka Sax}
\affiliation{Group of Applied Physics, University of Geneva, Chemin de Pinchat 22, CH-1211 Geneva 4, Switzerland}
\author{Alberto Boaron}
\affiliation{Group of Applied Physics, University of Geneva, Chemin de Pinchat 22, CH-1211 Geneva 4, Switzerland}
\author{Hugo Zbinden}
\affiliation{Group of Applied Physics, University of Geneva, Chemin de Pinchat 22, CH-1211 Geneva 4, Switzerland}

\begin{abstract}
Network integration of quantum key distribution is crucial for its future widespread deployment due to the high cost of using optical fibers dedicated for the quantum channel, only. We studied the performance of a system running a simplified BB84 protocol at \SI{2.5}{\giga\hertz} repetition rate, operating in the original wavelength band, short O-band, when multiplexed with communication channels in the conventional wavelength band, short C-band. Our system could successfully generate secret keys over a single-mode fiber with a length of \SI{95.5}{\kilo\meter} and with co-propagating classical signals at a launch power of \SI{8.9}{\dBm}. Further, we discuss the performance of an ideal system under the same conditions, showing the limits of what is possible with a discrete variable system in the O-band. We also considered a short and lossy link with \SI{51}{\kilo\meter} optical fiber resembling a real link in a metropolitan area network. In this scenario we could exchange a secret key with a launch power up to \SI{16.7}{\dBm} in the classical channels.
\end{abstract}

\maketitle


Quantum key distribution (QKD) allows for distribution of secret keys between distant parties. As of today, a variety of QKD experiments have shown the feasibility of exchanging keys through a dedicated optical fiber over hundreds of kilometers \cite{Boaron2018,Chen2020,Pittaluga2020}. However, the deployment and maintenance of optical fiber reserved for QKD only is rather costly and would hence limit the use cases of QKD. Therefore a pressing issue is the seamless integration of QKD into the already existing optical fiber network infrastructure. Using wavelength division multiplexing (WDM), it is possible to couple both QKD and classical communication signals to the same fiber \cite{Townsend1997}. The challenge of this approach lies in the fact that QKD protocols typically require a launch power of less than \SI{1}{\nano\watt}, whereas classical signals are launched with a power in the order of \SI{1}{\milli\watt} per channel. A small fraction of the classical signal arriving at the QKD receiver is enough to increase the quantum bit error rate (QBER) to a value where key extraction is impossible.

In many network environments, the classical signals populate the conventional wavelength band (C-band) from \SI{1530}{\nano\meter} to \SI{1565}{\nano\meter} separated by \SI{0.8}{\nano\meter} in a dense WDM (DWDM) grid. Upon coexisting with a quantum channel, a classical signal generates noise at the quantum receiver due to imperfect isolation between the DWDM channels or via non-linear processes. Raman scattering and, depending on the choice of the DWDM channels and the quantum channel wavelength, four-wave mixing are the dominant non-linear processes \cite{Eraerds2010,Aleksic2015, Tkach1995}.  While the channel isolation can be easily increased by adding suitable filters, non-linear processes can create photons at the same wavelength as the quantum signal which cannot be spectrally filtered. Four-wave mixing is restricted to narrow spectral regions and can therefore be avoided by choosing the quantum wavelength carefully. Raman noise, on the other hand, exhibits a broad spectrum. For example, classical signals in one C-band channel create a Raman noise spectrum covering the whole C-band with only two narrow local minima close to the pump wavelength \cite{Eraerds2010}. In a densely populated WDM environment, the local minima are covered by the Raman noise of other channels. 

One can make use of temporal filtering to help reduce the impact of noise photons at the quantum channel wavelength \cite{Patel2012,Mao2018}. The propagation direction of the classical signals also have an influence on the amount of introduced noise. A signal counter-propagating to the quantum signal introduces more Raman noise than a co-propagating one due to the isotropic nature of Raman scattering and the higher power in vicinity to the receiver \cite{Eraerds2010}.

Regarding the quantum channel wavelength, there are two frequent choices. Either it is placed in the C-band or in the original wavelength band (O-band) from \SI{1260}{\nano\meter} to \SI{1360}{\nano\meter}. The advantage of placing it in the C-band is the high fiber transmission. However, in a network, the quantum channel is then spectrally close to the classical channels and therefore strongly affected by Raman noise. Placing the quantum channel in the O-band reduces the amount of Raman noise but also the fiber transmission \cite{Townsend1997, Wang2017,Aleksic2015}. Generally speaking, it is advantageous to put the quantum signal in the O-band above a certain power threshold for the classical channels in the C-band \cite{Wang2017}. In present-day networks, the total loss of a link is often dominated by the excess loss due to fiber connections, routing devices or other components. In such an environment, a quantum channel in the O-band is advantageous since the transmission approaches the one of the C-band, but the noise is reduced. For both choices of quantum channel wavelength, the performance of QKD systems in the presence of classical communication has been studied \cite{Townsend1997, Mao2018, Frohlich2017,Kumar2015,Milovancev2021, Geng2021}. One study also considered a quantum channel in the long wavelength band (L-band) from \SI{1565}{\nano\meter} to \SI{1625}{\nano\meter} and the short wavelength band (S-band) from \SI{1460}{\nano\meter} to \SI{1530}{\nano\meter} \cite{Kleis2019}.

The performance of a QKD system in a network depends heavily on the quality of the noise filtering on the receiver side. First, high isolation of the quantum channel from the classical channel is needed. This can be easily achieved by cascading WDM modules. Second, high Raman noise rejection is desired. The quality of noise rejection depends on the time-bandwidth product of the quantum signal and on how tight the temporal and spectral filtering can be implemented. In the case of continuous variable (CV-)QKD systems, the homodyne detection acts as a spectral filter \cite{Qi2010}. For discrete variable (DV-)QKD systems, like the one presented in this study, filters have to be added at a cost of decreasing the transmission.

In this work we demonstrate the operation of a QKD system with a quantum channel in the  O-band with a wavelength of \SI{1310}{\nano\meter}. We consider a scenario where all the classical signals are co-propagating in the same fiber. This configuration is often found in metropolitan networks \cite{Ciurana2014, Aleksic2015}. The quantum channel is launched in the same direction as the classical channels to minimize the degradation of the quantum signal. We consider a channel where the loss is only given by the fiber attenuation and another channel where a substantial amount of loss is given by imperfections, which is a more realistic model for a network environment. Finally, we compare our setup to an ideal system in terms of temporal and spectral filtering.



\begin{figure*}
\includegraphics[width = 2\columnwidth]{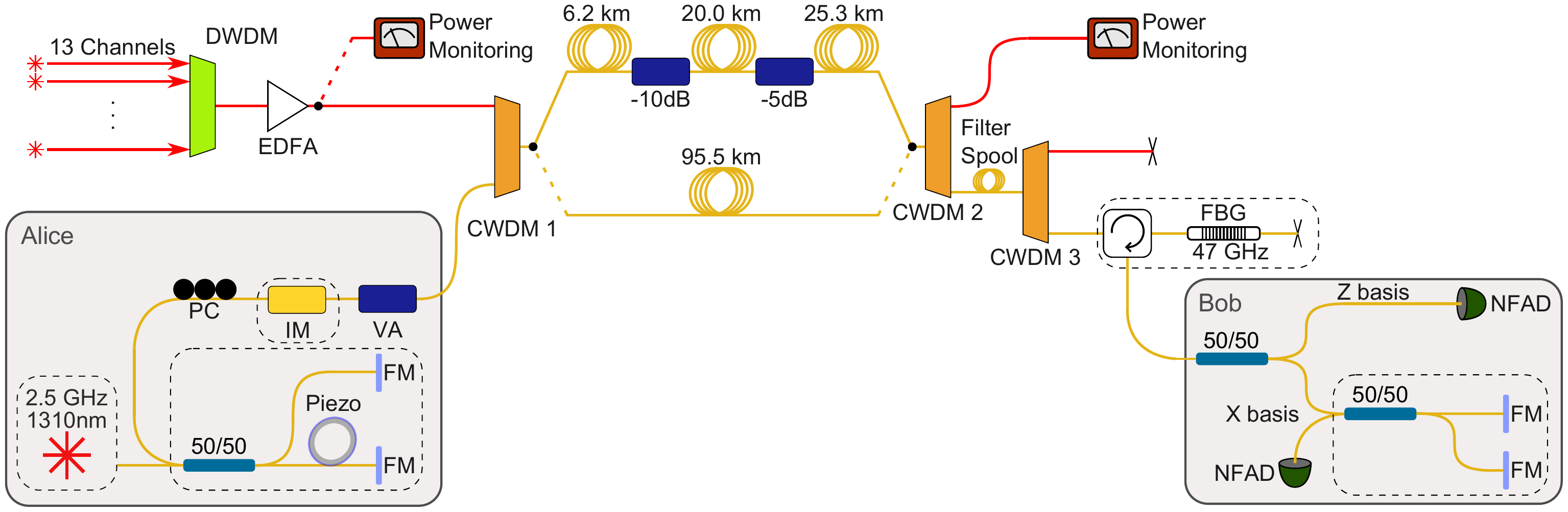}
\caption{\label{fig:setup} Schematic of the setup.  The dashed boxes are temperature stabilized. Fibers carrying the classical and quantum signals are shown in red and yellow, respectively. The transmission line is either a  \SI{95.5}{\kilo\meter} or a \SI{51.5}{\kilo\meter} long fiber, the latter is  intercepted by attenuators. The classical launch power was measured after the amplifier and the receiver power at the \SI{1550}{\nano\meter} port of the CWDM2.
EDFA: erbium doped fiber amplifier;
CWDM: coarse wavelength division multiplexer;
DWDM: dense wavelength division multiplexer;
FBG: fiber Bragg grating;
FM: Faraday mirror;
IM: intensity modulator;
NFAD: negative-feedback avalanche photodiode;
PC: polarization controller; 
VA: variable attenuator;
}
\end{figure*}

We utilize a simple time-bin protocol with one decoy state, operating at a qubit repetition rate of \SI{2.5}{\giga\hertz} \cite{Boaron2016}. The QKD implementation and the configuration of the classical channels are depicted in \autoref{fig:setup}. Alice encodes the qubits in two time-bins. The bits in the Z basis, which are used to generate the key, are encoded in either the early or late time bin. Only one state in the X basis is used, namely the superposition of the the early and late bin with a fixed relative phase. The laser source is a gain-switched distributed feedback laser emitting pulses with a full width at half maximum (FWHM) of \SI{26}{\pico\second}. Due to the gain-switching the pulses are chirped. 
  
Bob uses free-running InGaAs/InP negative-feedback avalanche diodes (NFADs) \cite{Amri2016}. Both NFADs are cooled to \SI{-85}{\celsius}, show a detection efficiency of 25\% at \SI{1310}{\nano\meter} and a jitter of \SI{50}{\pico\second}. The detector in the X basis (Z basis) shows a dark count rate of \SI{108}{\hertz} (\SI{91}{\hertz}).
 
The dead time was set to \SI{40}{\micro\second} for the detector in the X basis and to \SI{32}{\micro\second} for the one in the Z basis. The detection window per time bin has a duration of \SI{100}{\pico\second}. Detections outside this window are ignored by the acquisition system. The error correction was performed with a Cascade algorithm \cite{Martinez2015}, with an efficiency of 1.05. The compression factor was calculated over a privacy amplification block of \num{8e6} bits and taking into account finite-key effects \cite{Rusca2018b}.

The classical communication runs over thirteen C-band channels. They are multiplexed with a DWDM module and then amplified using an erbium-doped fiber amplifier. The quantum channel is added to the fiber with a coarse WDM (CWDM) module. On the receiver end, the quantum and classical signals are separated by a CWDM module. Another CWDM module is used to increase isolation between quantum and classical channels. To prevent classical signals to travel multiple times between the CWDM modules and to further improve the isolation, we added a fiber spool with a winding radius of \SI{16}{\milli\meter} and 36 windings. This spool has an insertion loss of \SI{1.0}{\deci\bel} at \SI{1310}{\nano\meter} and \SI{32.9}{\deci\bel} at \SI{1550}{\nano\meter}. The remaining signal and noise are filtered by a fiber Bragg grating (FBG) with a transmission window of \SI{47}{\giga\hertz} FWHM and more than \SI{30}{\deci\bel} of extinction outside the window.

\begin{table*}
\centering
\begin{tabular}{c | c | c | l } 
    \hline
    \thead{Description} & 
    \thead{Insertion loss \\ at \SI{1310}{\nano\meter} (dB)} &
    \thead{Isolation from \\ \SI{1550}{\nano\meter} (dB)} &  
    \thead{Remarks} \\
    \hline\hline 
 CWDM 1 & 0.8 & $> 45$ &  \\ \hline
 CWDM 2 & 0.6 & $> 45$ &  \\ \hline
 CWDM 3 & 0.8 & $> 45$ &  \\ \hline
 Filter Spool & 1.0 & 32.9 &  \\ \hline
\makecell{Fiber Bragg grating (FBG) \\ and circulator} & 
4.0 & $> 30$ &
\makecell{The insertion loss is partially caused by \\ 
spectral mismatch of laser pulse and filter. \\
A loss of \SI{1.8}{\deci\bel} was measured at peak transmission.} 
\\ \hline
\makecell{Loss due to detector jitter \\ and pulse broadening by FBG} & 
1.9 & - &
\makecell{The loss was obtained by observing the ratio \\
between detection events outside \\ 
and inside the detection time window.} \\ \hline
 \hline
\end{tabular}
\caption{Loss introduced by the filters. The filters are named the same as in \autoref{fig:setup}.}
\label{table:losses}
\end{table*}

The excess loss experienced by the quantum signal due to the filters is summarized in \autoref{table:losses}. The spectral width of the laser is close to the spectral width of the FBG, leading to increased insertion loss. Further, the FBG is slightly chirped and therefore the already chirped laser pulse gets temporally broadened by the FBG. The broadening due to the FBG together with the detector jitter increases the chance to detect the pulse outside the predefined time window, and therefore effectively introduces loss.


\begin{figure}
\includegraphics[ trim={0 0 0 0},clip, width = 1\columnwidth]{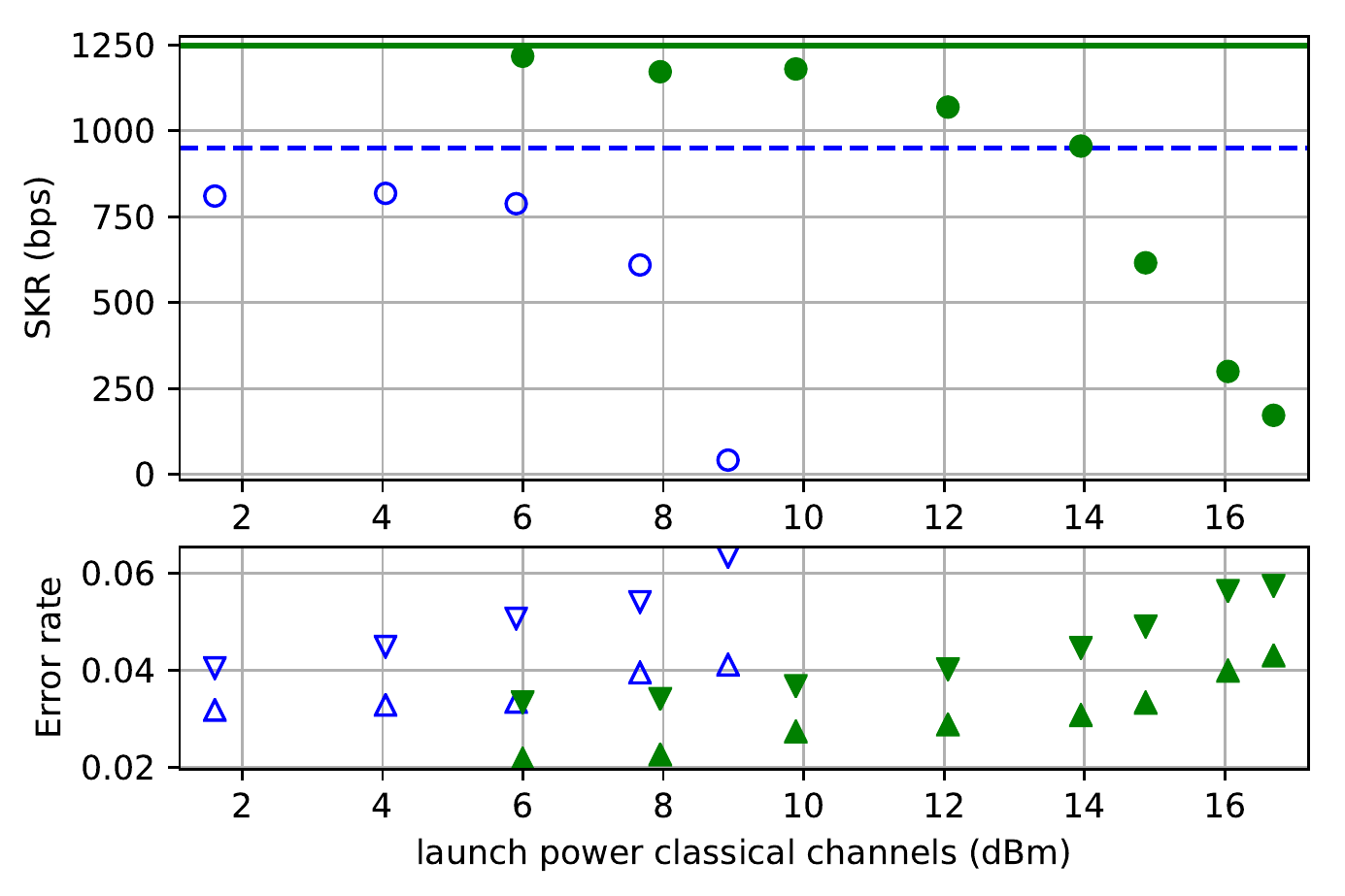}
\caption{\label{fig:results_95km} Measured secret key rate, phase error rate (triangles pointing up) and QBER Z (triangles pointing down) as a function of the total launch power in the classical channels. The filled green points were measured with the \SI{51.5}{\kilo\meter} link and the empty blue points were measured with the \SI{95.5}{\kilo\meter} link. The green solid line and the blue dashed line show the corresponding secret key rates without any classical signal.}
\end{figure}

We performed secret key exchanges in two different regimes. First, we considered a case similar to a real network where we used standard single mode fiber (Corning\textsuperscript{\textregistered} SMF-28e+\textsuperscript{\textregistered}) with a length of \SI{51.5}{\kilo\meter} together with \SI{15}{\dB} of excess loss in the channel (see \autoref{fig:setup} for details). This configuration acts as a model for a realistic link. In a metropolitan network, this loss could be due to connectors and routing equipment. Second, we exchanged a key over \SI{95.5}{\kilo\meter} of standard single-mode fiber (Corning\textsuperscript{\textregistered} SMF-28e+\textsuperscript{\textregistered}). This measurement was done for ease of comparison with previous studies. In both cases we are interested in the secret key rate as a function of the launch power in the classical channels.

In \autoref{fig:results_95km} we show the secret key rate, the QBER in the Z basis and the phase error rate as a function of the classical launch power for the two different channel configurations. We obtained a secret key rate of \SI{42}{\bps} with a launch power of \SI{8.9}{\dBm}, which corresponds to a total received power of \SI{-12.1}{\dBm}. In the case of the \SI{51.5}{\kilo\meter} long and high loss link, a secret key rate of \SI{172.2}{\bps} could be obtained at total launch power of \SI{16.7}{\dBm}, corresponding to a total received power of \SI{-11.8}{\dBm}.

We were also interested in finding the limits of what would be possible with an ideal setup using the same protocol, quantum channel wavelength and repetition rate as our experiment. For this we assume that the filter block on Bob's side (CWDM 3, filter spool, circulator and FBG in \autoref{fig:setup}) has negligible insertion loss, that the detectors have no jitter and no dark counts, that Alice is sending Fourier-limited $\text{sech}^2$-pulses and that the filter spectrum of the FBG would be optimized both in bandwidth and in shape for these pulses. Our simulation shows that in this case, the maximum tolerable launch power would increase by \SI{17.7}{\deci\bel}, where \SI{14.1}{\deci\bel} could be gained due to the absence of jitter, the optimized shape and bandwidth of the FBG and of the laser pulse and \SI{3.6}{\deci\bel} could be gained if we had an ideal filter block with negligible insertion loss. We also estimated that we the maximum tolerable launch power would increase by \SI{1.5}{\deci\bel} if we used superconducting nanowire single-photon detectors (SNSPDs) with a jitter of \SI{30}{\pico\second} instead of NFADs with a jitter of \SI{50}{\pico\second} as in our experimental setup.

\begin{table*}[t]
\centering
\begin{tabular}{c | c | c | c | c | c | c | c} 
    \hline
    \thead{continuous/discrete \\ variable QKD} & 
    \thead{Wavelength band} &  
    \thead{fiber length \\ (\si{\kilo\meter})} &  
    \thead{att. quantum \\ channel (\si{\deci\bel})} &  
    \thead{launch power \\ (\si{\dBm})} &  
    \thead{secret key \\ rate (bps)} &
    \thead{finite-key \\ statistics} &  
    \thead{Ref.}\\
    \hline\hline
\multirow{18}{*}{discrete}  & \multirow{12}{*}{O}  & 51.5 & 34.1 & 16.7 & \num{1.7e+02} & \multirow{4}{*}{Yes} & \multirow{4}{*}{This work} \\
 & & 51.5 & 34.1 & 13.9 & \num{9.6e+02} & \\
 & & 95.5 & 34.8 & 5.9 & \num{7.9e+02} & \\
 & & 95.5 & 34.8 & 8.9 & \num{4.2e+01} & \\
\cline{3-8}
  &   & 66.0 & 22.3 & 21.0 & \num{3.0e+03} & \multirow{3}{*}{Yes} & \multirow{3}{*}{\cite{Mao2018}} \\
 & & 66.0 & 22.3 & 16.0 & \num{3.9e+03} & \\
 & & 66.0 & 22.3 & 11.0 & \num{4.8e+03} & \\
\cline{3-8}
 &   & 40.0 & 12.8$^*$ & 17.6 & \num{5.0e+02} & \multirow{3}{*}{No} & \multirow{3}{*}{\cite{Geng2021}} \\
 & & 50.0 & 16.0$^*$ & 14.7 & \num{2.6e+02} & \\
 & & 60.0 & 19.2$^*$ & 11.7 & \num{1.8e+02} & \\
\cline{3-8}
  &  & 60.0 & 19.2$^*$ & 4.0 & \num{4.2e+03} & \multirow{2}{*}{Yes} & \multirow{2}{*}{\cite{Wang2017}} \\
 & & 80.0 & 25.6$^*$ & 4.0 & \num{1.2e+03} & \\
\cline{2-8}
  & \multirow{6}{*}{C}  & 50.0 & 9.6 & -18.5 & \num{1.1e+06} & \multirow{3}{*}{Yes} & \multirow{3}{*}{\cite{Dynes2016}} \\
 & & 50.0 & 9.6 & -12.5 & \num{8.6e+05} & \\
 & & 50.0 & 9.6 & -5.5 & \num{1.3e+05} & \\
\cline{3-8}
 &  & 100.0 & 18.0$^*$ & -3.1 & \num{1.0e+04} & \multirow{3}{*}{Yes} & \multirow{3}{*}{\cite{Frohlich2017}} \\
 & & 150.0 & 27.0$^*$ & -8.1 & \num{2.0e+03} & \\
 & & 150.0 & 27.0$^*$ & -5.0 & \num{1.2e+03} & \\
\hline
\multirow{7}{*}{continuous}  & \multirow{5}{*}{C}  & 25.0 & 5.0$^*$ & 11.5 & \num{1.2e+01} & \multirow{3}{*}{Yes} & \multirow{3}{*}{\cite{Kumar2015}} \\
 & & 75.0 & 15.0$^*$ & -0.5 & \num{9.0e+00} & \\
 & & 75.0 & 15.0$^*$ & -3.0 & \num{4.9e+02} & \\
\cline{3-8}
  & & 13.2 & 3.0$^*$ & 15.6 & \num{7.2e+07} & \multirow{2}{*}{No} & \multirow{2}{*}{\cite{Milovancev2021}} \\
 & & 28.4 & 6.4$^*$ & 15.6 & \num{2.8e+06} & \\
\cline{2-8}
  & \multirow{2}{*}{S}  & 25.0 & 5.0$^*$ & 14.0 & \num{4.0e+05} & \multirow{2}{*}{No} & \multirow{2}{*}{\cite{Kleis2019}} \\
 & & 50.0 & 10.0$^*$ & 6.0 & \num{1.7e+06} & \\
\hline
\hline
\end{tabular}
\caption{Comparison to previous studies. For each study, at least the points with the maximum fiber length and maximum launch power are mentioned. The classical channels are co-propagating with the quantum channel for all points shown here. The C-band  spans from \SI{1530}{\nano\meter} to \SI{1565}{\nano\meter}, the O-band from \SI{1260}{\nano\meter} to \SI{1360}{\nano\meter} and the S-band from \SI{1491}{\nano\meter} to \SI{1528}{\nano\meter}. $^*$The attenuation of the quantum channel was estimated from the fiber length.
}
\label{table:comparison}
\end{table*}

\begin{figure}
\begin{overpic}[scale=0.65,percent]{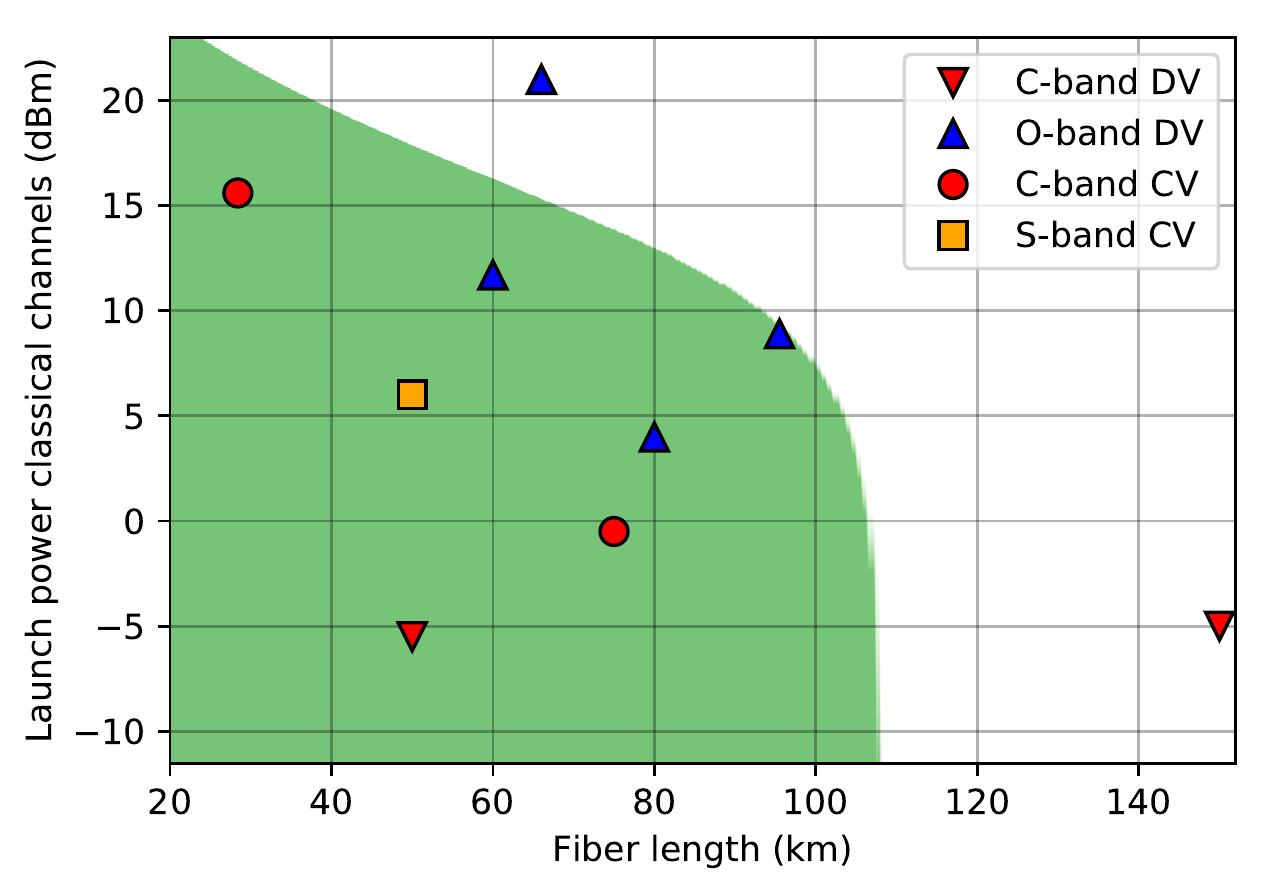}
{
\fontsize{8pt}{8pt}\selectfont
\put(54,46){This work}
\put(44,63){\cite{Mao2018} }
\put(89,20){\cite{Frohlich2017} }
\put(16,57){\cite{Milovancev2021} }
%
\put(30,42){\cite{Kleis2019} }
\put(53,35){\cite{Wang2017} }
%
\put(40.5,48){\cite{Geng2021} }
\put(51,28){\cite{Kumar2015} }
\put(30,23){\cite{Dynes2016} }
}
\end{overpic}%
\caption{\label{fig:comparison} Comparison to previous studies. The marker shape indicates if a continuous (CV) or discrete variable (DV) system  was used and in which Telecom band the quantum channel was situated. The green surface indicates the region where our experiment yields a positive secret key rate and was obtained by simulation with the same repetition rate, filter insertion loss, detector efficiency, pulse broadening, detector jitter and dark counts as in our setup. For all studies, the data point with the highest launch power in the classical channels at the highest fiber length is included in this figure.
}
\end{figure}


In \autoref{fig:comparison} and \autoref{table:comparison}, we compare our work to previous studies. In summary, CV-QKD systems show the best performance both in the tolerated launch power and in the secret key rate at short and low loss links. At longer distances DV-QKD systems, both in the C- and O-band, outperform the CV-QKD systems. We can conclude that, as of today, DV-QKD systems operating in the O-band are best suited for networks with distances between \SI{50}{\kilo\meter} and \SI{95}{\kilo\meter} and high launch power. Furthermore, our results in \autoref{fig:results_95km} with the short and high loss link show that in a real network, O-band DV-QKD systems can tolerate more power than suggested by measurements with links where the loss is mainly given by the fiber. Another advantage of O-band QKD systems is that separating a signal from the C-band is possible with rather low loss and high isolation by using CWDM modules. In our case one CWDM module has an insertion loss of \SI{0.8}{\deci\bel} and a channel isolation of more than \SI{45}{\deci\bel}. If we wanted to isolate one C-band channel from other C-band channels we would need DWDM modules, which typically exhibit an insertion loss of \SI{2.5}{\deci\bel} while having a channel isolation of only \SI{25}{\deci\bel}. Therefore it is best to use O-band QKD systems in metropolitan area networks.

In conclusion, we showed that with a simple and practical QKD system it is possible to exchange a secret key in presence of classical channels. We demonstrated the feasibility of key generation for a short distance and high loss link and also for a medium range link where the loss is predominantly given by the fiber attenuation. An ideal DV-QKD system at a repetition rate of \SI{2.5}{\giga\hertz} in the O-band could tolerate a total launch power of \SI{27}{\dBm} of co-propagating classical signals over \SI{95}{\kilo\meter} of single-mode fiber. This would be even enough to operate the QKD system in a backbone fiber network \cite{Mao2018}. Finding ways to prepare almost ideal pulses and manufacturing optimized filters could increase noise tolerance of DV-QKD systems by more than an order of magnitude according to our simulation.



We thank Romain Alléaume and Eleni Diamanti for the useful discussions and the Swiss NCCR QSIT for financial support.


The following article has been accepted by Applied Physics Letters. After it is published, it will be found at \href{https://doi.org/10.1063/5.0060232}{https://doi.org/10.1063/5.0060232}.

\section*{Data availability}

The data that support the findings of this study are available from the corresponding author upon reasonable request.

\bibliography{bibliography}
\end{document}